\documentclass{ws-procs9x6}

\begin{document}

\title{Texture Zeros and CP-violating Phases \\
in the Neutrino Mass Matrix}
\author{Zhi-zhong Xing}
\address{Institute of High Energy Physics, Chinese Academy of Sciences \\
P.O. Box 918 (4), Beijing 100039, China \\
E-mail: xingzz@mail.ihep.ac.cn}

\maketitle

\abstracts{We stress that specific texture zeros of lepton mass
matrices, which might dynamically arise from a new kind of flavor
symmetry, can help us to establish simple and testable relations
between the lepton flavor mixing parameters and lepton mass
ratios. We present a brief review of one-zero, two-zero and
three-zero textures of the neutrino mass matrix. Their
phenomenological consequences on neutrino mixing and CP violation
are also discussed.}

\section{Introduction}

Impressively robust evidence in favor of neutrino oscillations has
been accumulated from the solar\cite{SNO}, atmospheric\cite{SK},
reactor\cite{KM} and accelerator\cite{K2K} neutrino experiments in
the past few years. We are now convinced that neutrinos are
massive and lepton flavors are mixed. In spite of such exciting
progress in neutrino physics, our quantitative knowledge about
neutrino masses and lepton flavor mixing remain rather poor -- for
example, the absolute scale of three neutrino masses, the smallest
lepton mixing angle and three CP-violating phases are still
unknown.

In the lack of a convincing flavor theory, five approaches have
been tried towards a deeper understanding of fermion mass
generation\cite{X04}: (a) radiative mechanisms\cite{a}; (b)
texture zeros\cite{b}; (c) flavor symmetries\cite{c}; (d) seesaw
mechanisms\cite{d}; and (e) extra dimensions\cite{e}. Some of them
can be correlated with one another. For instance, lepton mass
matrices may have a few texture zeros as a natural consequence of
a new kind of flavor symmetry\cite{Tanimoto04}, and those texture
zeros may guarantee some calculability and allow us to predict the
neutrino mass spectrum and lepton flavor mixing parameters via the
seesaw mechanism. Note that texture zeros of a fermion mass matrix
dynamically mean that the corresponding matrix elements are
sufficiently suppressed in comparison with their neighboring
counterparts. A very good lesson from the quark sector is
remarkable: reasonable zeros of quark mass matrices allow us to
establish some simple and testable relations between the flavor
mixing angles and quark mass ratios\cite{FX} -- if such relations
are more or less unique and experimentally favored, they may have
a good chance to be close to the truth -- namely, the same or
similar relations should be predicted by the underlying (true)
theory with much fewer fundamental parameters. Hence a
phenomenological study of possible texture zeros of fermion mass
matrices {\it does} make some sense to get useful hints about
flavor dynamics responsible for the generation of fermion masses
and the origin of CP violation.

The phenomenology of lepton masses and flavor mixing at low
energies can be formulated in terms of the charged lepton mass
matrix $M_l$ and the (effective) neutrino mass matrix $M_\nu$. The
lepton flavor mixing matrix $V$ arises from the mismatch between
diagonalizations of $M_l$ and $M_\nu$. There are totally twelve
physical parameters: three charged lepton masses $(m_e, m_\mu,
m_\tau)$, three neutrino masses $(m_1, m_2, m_3$), three flavor
mixing angles $(\theta_{12}, \theta_{23}, \theta_{13})$, and three
CP-violating phases $(\delta, \rho, \sigma)$. Besides $m_e$,
$m_\mu$ and $m_\tau$, preliminary values of $\Delta m^2_{21}$,
$|\Delta m^2_{32}|$, $\theta_{12}$ and $\theta_{23}$ have
essentially been extracted from solar and atmospheric neutrino
oscillations. How small $\theta_{13}$ is remains an open
question\cite{Mei}. The sign of $\Delta m^2_{32}$ is unknown and
the magnitudes of $\delta$, $\rho$ and $\sigma$ are entirely
unrestricted. It seems impossible to fully determine $M_l$ and
$M_\nu$ from the feasible experiments at present or in the near
future. In this situation, we hope that possible texture zeros of
$M_l$ and $M_\nu$ may help us out.

We remark that texture zeros of lepton mass matrices can lead to
some simple and testable relations between unknown and known
parameters of neutrino oscillations. Of course, such zeros may not
be preserved to all orders or at any energy scales in the
unspecified interactions from which lepton masses are generated.
At the one-loop level and in the flavor basis where $M_l$ is
diagonal and positive, however, the renormalization-group
evolution of $M_\nu$ from the seesaw scale (i.e., the mass scale
of the lightest right-handed Majorana neutrino) to the electroweak
scale {\it does} allow its texture zeros to preserve\cite{FX}.
Once the approach of texture zeros is combined with the seesaw
mechanism, it is possible to simultaneously account for the
cosmological baryon number asymmetry via leptogenesis\cite{FY} and
current neutrino oscillation data.

This talk is subject to a phenomenological analysis of texture
zeros of lepton mass matrices at low energy scales. For
simplicity, we restrict ourselves to symmetric $M_l$ and $M_\nu$.
A symmetric lepton mass matrix totally has six independent
entries. If $n$ of them is/are taken to be vanishing, we will
arrive at
$$
^6{\bf C}_n \; =\; \frac{6!}{n! \left (6 - n\right )!}
$$
patterns, which are structurally different from one another. It is
obvious that a pattern of $M_l$ or $M_\nu$ with more than three
texture zeros (i.e., $n\geq 4$) has no chance to be compatible
with the experimental data of lepton masses and flavor mixing
angles. Hence we shall pay our attention to 20 three-zero
textures, 15 two-zero textures and 6 one-zero textures of $M_\nu$
in the following.

\section{Three-zero textures of $M_\nu$}

There are twenty three-zero patterns of $M_l$ or $M_\nu$, which
can be classified into four categories:

\hspace{-0.7cm} (a) Three diagonal matrix elements are all
vanishing (type 0):
$$
M_0 = \left ( \begin{array}{ccc} {\bf 0} & \times    & \times \\
\times & {\bf 0} & \times \\ \times  & \times    & {\bf 0}
\end{array} \right ) \; , ~~~~~~~
$$
where those non-vanishing entries are simply symbolized by
$\times$'s.

\hspace{-0.7cm} (b) Two diagonal matrix elements are vanishing
(type I):
\begin{eqnarray}
&& M_{\rm I_1} = \left ( \begin{array}{ccc} {\bf 0} & \times &
{\bf 0} \\ \times  & {\bf 0} & \times \\ {\bf 0} & \times  &
\times \end{array} \right ) \; , ~~~ M_{\rm I_2} = \left (
\begin{array}{ccc} {\bf 0} & {\bf 0} & \times \\ {\bf 0} & \times  &
\times \\ \times  & \times  & {\bf 0} \end{array} \right ) \; ,
~~~ M_{\rm I_3} = \left ( \begin{array}{ccc} {\bf 0} & \times &
\times \\ \times  & {\bf 0} & {\bf 0} \\ \times & {\bf 0} & \times
\end{array} \right ) \; , ~~~~~~
\nonumber \\
&& M_{\rm I_4} = \left ( \begin{array}{ccc} {\bf 0} & \times  &
\times \\ \times & \times  & {\bf 0} \\ \times  & {\bf 0} & {\bf
0} \end{array} \right ) \; , ~~~ M_{\rm I_5} = \left (
\begin{array}{ccc} \times & {\bf 0} & \times \\ {\bf 0} & {\bf 0} &
\times \\ \times  & \times  & {\bf 0} \end{array} \right ) \; ,
~~~ M_{\rm I_6} = \left ( \begin{array}{ccc} \times  & \times  &
{\bf 0} \\ \times & {\bf 0} & \times  \\ {\bf 0} & \times  & {\bf
0} \end{array} \right ) \; , ~~~~~~ \nonumber
\end{eqnarray}
which are of rank three; and
$$
~~~ M_{\rm I_7} = \left ( \begin{array}{ccc} {\bf 0} & {\bf 0} &
\times \\ {\bf 0} & {\bf 0} & \times \\ \times  & \times  & \times
\end{array} \right ) \; , ~~~ M_{\rm I_8} = \left ( \begin{array}{ccc} {\bf 0} &
\times & {\bf 0} \\ \times  & \times  & \times \\ {\bf 0} & \times
& {\bf 0} \end{array} \right ) \; , ~~~ M_{\rm I_9} = \left (
\begin{array}{ccc} \times  & \times  & \times \\ \times  & {\bf 0} &
{\bf 0} \\ \times  & {\bf 0} & {\bf 0} \end{array} \right ) \; ,
~~~~~~
$$
which are of rank two.

\hspace{-0.7cm} (c) One diagonal matrix element is vanishing (type
II):
\begin{eqnarray}
&& M_{\rm II_1} = \left ( \begin{array}{ccc} \times  & \times  &
{\bf 0} \\ \times & {\bf 0} & {\bf 0} \\ {\bf 0} & {\bf 0} &
\times \end{array} \right ) \; , ~~~ M_{\rm II_2} = \left (
\begin{array}{ccc} \times & {\bf 0} & \times \\ {\bf 0} & \times  &
{\bf 0} \\ \times  & {\bf 0} & {\bf 0} \end{array} \right ) \; ,
~~~ M_{\rm II_3} = \left ( \begin{array}{ccc} {\bf 0} & \times &
{\bf 0} \\ \times  & \times & {\bf 0} \\ {\bf 0} & {\bf 0} &
\times \end{array} \right ) \; , ~~~~~
\nonumber \\
&& M_{\rm II_4} = \left ( \begin{array}{ccc} {\bf 0} & {\bf 0} &
\times \\ {\bf 0} & \times  & {\bf 0} \\ \times  & {\bf 0} &
\times \end{array} \right ) \; , ~~~ M_{\rm II_5} = \left (
\begin{array}{ccc} \times & {\bf 0} & {\bf 0} \\ {\bf 0} & \times  & \times
\\ {\bf 0} & \times & {\bf 0} \end{array} \right ) \; , ~~~ M_{\rm II_6} =
\left ( \begin{array}{ccc} \times & {\bf 0} & {\bf 0} \\ {\bf 0} &
{\bf 0} & \times \\ {\bf 0} & \times  & \times  \end{array} \right
) \; , ~~~~~ \nonumber
\end{eqnarray}
which are of rank three; and
$$
~~~ M_{\rm II_7} = \left ( \begin{array}{ccc} \times  & \times  &
{\bf 0}
\\ \times  & \times  & {\bf 0} \\ {\bf 0} & {\bf 0} & {\bf 0} \end{array}
\right ) \; , ~~~ M_{\rm II_8} = \left ( \begin{array}{ccc} \times
& {\bf 0} & \times \\ {\bf 0} & {\bf 0} & {\bf 0} \\ \times & {\bf
0} & \times  \end{array} \right ) \; , ~~~ M_{\rm II_9} = \left (
\begin{array}{ccc} {\bf 0} & {\bf 0} & {\bf 0} \\ {\bf 0} & \times
& \times \\ {\bf 0} & \times  & \times  \end{array} \right ) \; ,
~~~~~
$$
which are of rank two.

\hspace{-0.7cm} (d) Three diagonal matrix elements are all
non-vanishing (type III):
$$
M_{\rm III} = \left ( \begin{array}{ccc} \times  & {\bf 0} & {\bf
0} \\ {\bf 0} & \times  & {\bf 0} \\ {\bf 0} & {\bf 0} & \times
\end{array} \right ) \; . ~~~~~~~
$$
This pattern itself does not give rise to flavor mixing.

In the flavor basis where $M_l$ is diagonal and positive (i.e., of
the pattern $M_{\rm III}$) and $M_\nu$ takes one of the above 20
patterns, we find that none of the 20 combinations of $M_l$ and
$M_\nu$ can be compatible with current neutrino oscillation data.
When $M_l$ is allowed to have off-diagonal non-vanishing entries,
however, the situation will change. A careful analysis\cite{ZX}
shows that there are totally 24 combinations of $M_l$ and $M_\nu$
with six texture zeros, which are compatible with current
experimental data at the $3\sigma$ level. These 24 patterns can be
classified into a few distinct categories:
\begin{eqnarray}
&&
\begin{tabular}{|c|c|c|c|c|c|c|} \hline
~ $M_l$ ~ & ~ $\rm I_1$ ~ & ~ $\rm I_2$ ~ & ~ $\rm I_3$ ~ & ~ $\rm
I_4$ ~ & ~ $\rm I_5$ ~ & ~ $\rm I_6$ ~ \\ \hline
~ $M_\nu$ ~ & ~ $\rm I_1$ ~ & ~ $\rm I_2$ ~ & ~ $\rm I_3$ ~ & ~
$\rm I_4$ ~ & ~ $\rm I_5$ ~ & ~ $\rm I_6$ ~ \\ \hline
\end{tabular} \;\;\; ,
\nonumber \\
&&
\begin{tabular}{|c|c|c|c|c|c|c|} \hline
~ $M_l$ ~ & ~ $\rm I_1$ ~ & ~ $\rm I_2$ ~ & ~ $\rm I_3$ ~ & ~ $\rm
I_4$ ~ & ~ $\rm I_5$ ~ & ~ $\rm I_6$ ~ \\ \hline
~ $M_\nu$ ~ & ~ $\rm I_2$ ~ & ~ $\rm I_1$ ~ & ~ $\rm I_5$ ~ & ~
$\rm I_6$ ~ & ~ $\rm I_3$ ~ & ~ $\rm I_4$ ~ \\ \hline
\end{tabular} \;\;\; ,
\nonumber
\end{eqnarray}
and
\begin{eqnarray}
&&
\begin{tabular}{|c|c|c|c|} \hline
~ $M_l$ ~ & ~ $\rm I_1$ ~ & ~ $\rm I_4$ ~ & ~ $\rm I_5$ ~ \\
\hline
~ $M_\nu$ ~ & ~ $\rm I_7$ ~ & ~ $\rm I_8$ ~ & ~ $\rm I_9$ ~ \\
\hline
\end{tabular} \;\;\; ,
~~~
\begin{tabular}{|c|c|c|c|} \hline
~ $M_l$ ~ & ~ $\rm I_3$ ~ & ~ $\rm I_2$ ~ & ~ $\rm I_6$ ~ \\
\hline
~ $M_\nu$ ~ & ~ $\rm I_7$ ~ & ~ $\rm I_8$ ~ & ~ $\rm I_9$ ~ \\
\hline
\end{tabular} \;\;\; ,
\nonumber \\
&&
\begin{tabular}{|c|c|c|c|} \hline
~ $M_l$ ~ & ~ $\rm I_2$ ~ & ~ $\rm I_6$ ~ & ~ $\rm I_3$ ~ \\
\hline
~ $M_\nu$ ~ & ~ $\rm I_7$ ~ & ~ $\rm I_8$ ~ & ~ $\rm I_9$ ~ \\
\hline
\end{tabular} \;\;\; ,
~~~
\begin{tabular}{|c|c|c|c|} \hline
~ $M_l$ ~ & ~ $\rm I_5$ ~ & ~ $\rm I_1$ ~ & ~ $\rm I_4$ ~ \\
\hline
~ $M_\nu$ ~ & ~ $\rm I_7$ ~ & ~ $\rm I_8$ ~ & ~ $\rm I_9$ ~ \\
\hline
\end{tabular} \;\;\; .
\nonumber
\end{eqnarray}
The lepton mass matrices in each category are isomeric -- namely,
they have different structures but their phenomenological
consequences are exactly the same \cite{ZX}. Note that $m_1 =0$
holds for the last four categories of lepton mass matrices, in
which $M_\nu$ is of rank two. Both the neutrino mass spectrum
($m_1$, $m_2$ and $m_3$) and the Majorana phases of CP violation
($\rho$ and $\sigma$) can well be determined or constrained for
these 24 patterns.

\section{Two-zero textures of $M_\nu$}

In the past two years, some particular attention has been paid to
two-zero textures of the neutrino mass matrix in the flavor basis
where the charged lepton mass matrix is diagonal and
positive\cite{Glashow,Xing02}. There are totally fifteen possible
patterns of $M_\nu$ with two independent vanishing entries, as
illustrated below.
\begin{center}
\begin{tabular}{c|c|c} \hline\hline
~~~~ Pattern $\rm A_1$ ~~~~&~~~~ Pattern $\rm A_2$ ~~~~&~~~~
Pattern $\rm B_1$ ~~~~ \\ \hline $\left ( \begin{array}{ccc} {\bf
0} & {\bf 0} & \times \\ {\bf 0} & \times & \times \\ \times &
\times & \times \end{array} \right )$ & $\left (
\begin{array}{ccc} {\bf 0} & \times & {\bf 0} \cr \times & \times
& \times \cr {\bf 0} & \times & \times \end{array} \right )$ &
$\left ( \begin{array}{ccc} \times & \times & {\bf 0} \cr \times &
{\bf 0} & \times \cr {\bf 0} & \times & \times \end{array} \right )$ \\
\hline\hline ~~~~ Pattern $\rm B_2$ ~~~~&~~~~ Pattern $\rm B_3$
~~~~&~~~~ Pattern $\rm B_4$ ~~~~
\\ \hline $\left ( \begin{array}{ccc} \times & {\bf 0} & \times \cr
{\bf 0} & \times & \times \cr \times & \times & {\bf 0}
\end{array} \right )$ & $\left ( \begin{array}{ccc} \times & {\bf 0}
& \times \cr {\bf 0} & {\bf 0} & \times \cr \times & \times &
\times \end{array} \right )$ & $\left ( \begin{array}{ccc} \times
& \times & {\bf 0} \cr \times &
\times & \times \cr {\bf 0} & \times & {\bf 0} \end{array} \right )$ \\
\hline\hline ~~~~ Pattern $\rm C$ ~~~~&~~~~ Pattern $\rm D_1$
~~~~&~~~~ Pattern $\rm D_2$ ~~~~ \\
\hline $\left ( \begin{array}{ccc} \times & \times & \times \cr
\times & {\bf 0} & \times \cr \times & \times & {\bf 0}
\end{array} \right )$ & $\left ( \begin{array}{ccc} \times & \times
& \times \cr \times & {\bf 0} & {\bf 0} \cr \times & {\bf 0} &
\times \end{array} \right )$ & $\left ( \begin{array}{ccc} \times
& \times & \times \cr \times &
\times & {\bf 0} \cr \times & \times & {\bf 0} \end{array} \right )$ \\
\hline\hline ~~~~ Pattern $\rm E_1$ ~~~~&~~~~ Pattern $\rm E_2$
~~~~&~~~~ Pattern $\rm E_3$ ~~~~ \\ \hline $\left (
\begin{array}{ccc} {\bf 0} & \times & \times \cr \times & {\bf 0} &
\times \cr \times & \times & \times \end{array} \right )$ & $\left
( \begin{array}{ccc} {\bf 0} & \times & \times \cr \times & \times
& \times \cr \times & \times & {\bf 0} \end{array} \right )$ &
$\left ( \begin{array}{ccc} {\bf 0} & \times & \times \cr \times &
\times & {\bf 0} \cr \times & {\bf 0} & \times \end{array} \right )$ \\
\hline\hline ~~~~ Pattern $\rm F_1$ ~~~~&~~~~ Pattern $\rm F_2$
~~~~&~~~~ Pattern $\rm F_3$ ~~~~ \\ \hline $\left (
\begin{array}{ccc} \times & {\bf 0} & {\bf 0} \cr {\bf 0} & \times &
\times \cr {\bf 0} & \times & \times \end{array} \right )$ &
$\left ( \begin{array}{ccc} \times & {\bf 0} & \times \cr {\bf 0}
& \times & {\bf 0} \cr \times & {\bf 0} & \times \end{array}
\right )$ & $\left ( \begin{array}{ccc} \times & \times & {\bf 0}
\cr \times &
\times & {\bf 0} \cr {\bf 0} & {\bf 0} & \times \end{array} \right )$ \\
\hline\hline
\end{tabular}
\end{center}
Among these patterns, seven of them ($\rm A_{1,2}$, $\rm
B_{1,2,3,4}$ and $\rm C$) are found to be compatible with current
neutrino oscillation data; and two of them ($\rm D_{1,2}$) are
only marginally allowed by today's experimental data. The left six
patterns ($\rm E_{1,2,3}$ and $\rm F_{1,2,3}$) have been ruled out
in phenomenology.

One may reproduce those phenomenologically-favored two-zero
textures of $M_\nu$ with the help of the seesaw mechanism and
calculate the cosmological baryon number asymmetry via
leptogenesis\cite{Tanimoto}.

\section{One-zero textures of $M_\nu$}

There are totally six different textures of the neutrino mass
matrix with one independent vanishing entry, as shown below.
\begin{center}
\begin{tabular}{c|c|c} \hline\hline
~~~~ Pattern $\rm A$ ~~~~&~~~~ Pattern $\rm B$ ~~~~&~~~~ Pattern
$\rm C$ ~~~~ \\ \hline $\left ( \begin{array}{ccc} {\bf 0} &
\times & \times \\ \times & \times & \times \\ \times & \times &
\times \end{array} \right )$ & $\left (
\begin{array}{ccc} \times & {\bf 0} & \times \cr {\bf 0} & \times
& \times \cr \times & \times & \times \end{array} \right )$ &
$\left ( \begin{array}{ccc} \times & \times & {\bf 0} \cr \times &
\times & \times \cr {\bf 0} & \times & \times \end{array} \right )$ \\
\hline\hline ~~~~ Pattern $\rm D$ ~~~~&~~~~ Pattern $\rm E$
~~~~&~~~~ Pattern $\rm F$ ~~~~
\\ \hline $\left ( \begin{array}{ccc} \times & \times & \times \cr
\times & {\bf 0} & \times \cr \times & \times & \times
\end{array} \right )$ & $\left ( \begin{array}{ccc} \times &
\times & \times \cr \times & \times & {\bf 0} \cr \times & {\bf 0}
& \times \end{array} \right )$ & $\left (
\begin{array}{ccc} \times & \times & \times \cr \times &
\times & \times \cr \times & \times & {\bf 0} \end{array} \right )$ \\
\hline\hline
\end{tabular}
\end{center}
In the flavor basis where $M_l$ is diagonal and positive, one may
confront these one-zero textures of $M_\nu$ with current neutrino
oscillation data. However, it is impossible to fully determine the
neutrino mass spectrum and Majorana phases of CP violation in this
case.

Note that pattern A is of particular interest, because it predicts
$\langle m\rangle_{ee} = 0$ (namely, the effective mass of the
neutrinoless double beta decay vanishes). While $\langle
m\rangle_{ee} \neq 0$ must imply that neutrinos are Majorana
particles, $\langle m\rangle_{ee} = 0$ does not {\it necessarily}
imply that neutrinos are Dirac particles. The reason is simply
that the vanishing or suppression of $\langle m\rangle_{ee}$ may
be due to large cancellation induced by the Majorana CP-violating
phases. Current neutrino oscillation data together with the
condition $\langle m\rangle_{ee} = 0$ can well constrain the
neutrino mass ratios and two Majorana phases\cite{Xing03}.

Now we consider one-zero textures of $M_\nu$ with one vanishing
eigenvalue ($m_1 =0$ or $m_3 =0$). Such phenomenological scenarios
are interesting, because they may naturally appear in the minimal
seesaw model with two heavy right-handed neutrinos\cite{FGY}. A
careful analysis\cite{Xing04} shows that patterns A, B and C with
$m_1 =0$ are compatible with current neutrino data, so are
patterns B, C, D and F with $m_3 =0$. In particular, the
CP-violating phases are calculable in the minimal seesaw model
with the Frampton-Glashow-Yanagida-like
ans$\rm\ddot{a}$tze\cite{GX}. Radiative corrections to neutrino
masses and lepton flavor mixing parameters have also been computed
in this framework\cite{MX}.

\section{Concluding remarks}

We have briefly described one-zero, two-zero and three-zero
textures of the neutrino mass matrix and confronted them with
today's neutrino oscillation data. Some more remarks and comments
are in order.

(a) Vanishing entries of simplified $M_l$ and $M_\nu$ (or the
flavor mixing matrix $V$) may serve as a symmetry limit of small
entries of realistic $M_l$ and $M_\nu$ (or $V$). The latter may
come from explicit perturbations or radiative corrections. Model
building may start from the possible symmetry limit, in which CP
might be conserving. For example, the bi-large mixing pattern of
$V$ with small $|V_{e3}|$ at low energies might result from the
bi-maximal mixing pattern of $V$ with vanishing $|V_{e3}|$ at GUT
scales. In this case, the seesaw threshold effects play a primary
role\cite{Mei,Ratz}.

(b) Lepton and quark mass matrices could have the same texture
zeros, arising from a universal flavor symmetry at a certain
energy scale. For instance, a typical four-zero texture of the
form
\begin{equation}
M \; \sim \; \left ( \begin{array}{ccc} {\bf 0} & \times & {\bf 0}
\cr \times & \times & \times \cr {\bf 0} & \times & \times
\end{array} \right ) \;
\end{equation}
is favored for quark mass matrices and is
seesaw-invariant\cite{FX}. It can be incorporated into a
SO(10)-inspired neutrino model\cite{BW}.

(c) It will be very useful to explore the full parameter space of
$M_\nu$ by using more accurate experimental data in no assumption
of texture zeros\cite{FS}. Such an analysis might more or less
favor certain of texture zeros in $M_\nu$, nevertheless.

Our conclusion is that the mass hierarchy of charged leptons and
quarks (and perhaps neutrinos) seem to imply certain textures of
fermion mass matrices, in which zeros may (approximately) be
present. The study of texture zeros could help establish the true
bridge between experimental measurables and fundamental parameters
of the underlying flavor dynamics.

\section*{Acknowledgments}
This work was supported in part by the National Natural Science
Foundation of China.

\end{document}